\documentstyle[sprocl,epsf]{article}

\def\be{\begin{equation}}
\def\ee{\end{equation}}
\def\bea{\begin{eqnarray}}
\def\eea{\end{eqnarray}}

\def\openone{\leavevmode\hbox{\small1\kern-3.8pt\normalsize1}}%

\def\lsim{{~\raise.15em\hbox{$<$}\kern-.85em
          \lower.35em\hbox{$\sim$}~}}

\newlength{\miniwidth}
\newlength{\miniwidthplot}
\newlength{\nseparation}
\newenvironment{nfigure}[1]
        {\begin{figure}[#1]\hrule\vspace{\nseparation}\par}
        {\vspace{\nseparation}\par \hrule \end{figure}}

\begin{document}

\title{NEW PHYSICS SEARCH IN B MESON DECAYS}

\author{Y.-Y. KEUM}

\address{Institute of Physics, Academia Sinica, Nankang, Taipei, 
Republic of China
\\E-mail: keum@phys.sinica.edu.tw} 

\maketitle\abstracts{ 
We discuss new physics search  
within the minimal supergravity model
by using a possible
large direct CP asymmetry in $B^{\pm} \to K^{\pm} \phi, K^{0}\pi^{\pm}$
decays and B meson rare decays.}

\section{Introduction}
$B$ fatories at KEK and SLAC are taking data to probe the origin of CP
violation which is one of main issue in current high energy physics.
In the standard model(SM) the CP violation is originated 
by a physical  phase of
the Cabbibo-Kobayashi-Maskawa(CKM) matrix \cite{ckm}.
A new source of CP violation can appear in models beyond the SM.
If there is new physics beyond SM, we expect to see  its effects in CP
violating B meson decays.

The most important task for new physics search 
is to identify decay modes 
where one can find a large deviations from standard model 
expectations, and also experimentally accessible in near future. 
It is really good chance to probe 
the new physics effects in B-meson decays through indirect 
($B^0-\bar{B}^0$ mixing) 
and direct CP-violating processes at B factories.

In this talk we discuss
a possible new physics impact on B-meson rare decays and 
the direct CP violation
phenomena through the magetic gluon penguin contributions
in $B^{\pm} \to K^{\pm}\phi, K^{0}\pi^{\pm}$ decays.

\section{New physics Effects in mSUGRA Model}
We investigate the new physics effects in rare B decays:
$B \to X_s \gamma$ and $B \to X_s \ell^{+} \ell^{-}$ and 
in direct CP violating modes: $B^{\pm} \to \phi K^{\pm}, K^{0} \pi^{\pm}$
within the minimal Supergravity model(mSUGRA).
In the mSUGRA, there are four new CP violating phases, i.e. 
phases of the gaugino mass, the higgsino mass parameter, the SUSY
breaking Higgs beson mass, and the trilinear scalar coupling constant,
of which two combinations are physically independent.
When we impose the universal condition at GUT scale, two physical
phases at GUT scale comes from, if we take a phase convention,
the trilinear coupling constant and higgsino mass parameter,
$\phi_{A}$ and $\phi_{\mu}$, respectively. 
These phases induce the neutron and electron electric dipole moments
(EDMs). When we require the unversality of SUSY breaking term at GUT scale
and explicitly solve the renormalization group equations (RGEs) 
to determine the masses and the mixings of SUSY particles and also
require the condition for the radiative electroweak breaking,
the phase $\phi_{\mu}$ is strongly constrained by EDMs and the phase
$\phi_{A}$ is not constrainted at GUT scale, however, in low energy scale,
the phase of A-term for top squarks is strongly supressed, becuase
the phase of A-term for top squarks is reduced due to the large top
Yukawa coupling constant and aligned to that of the gaugino mass \cite{kek}.

When we investigate the effect of the SUSY CP violating parameter
on rare B decays, $B \to X_s \gamma$ and $B \to X_s \ell^{+}\ell^{-}$,
in the mSUGRA, we find some interesting results \cite{kek} with following
SUSY parameters: $0< m_0 < 1 $ TeV, $120 < M_X < 500$ GeV, $|A_x| < 5 m_0$,
and the bound of EDMs\cite{edm} :
$|d_n| \leq 0.97 \times 10^{-25} e \cdot cm$, 
$|d_e| \leq 4.0 \times 10^{-27} e \cdot cm$, in addition, the branching ratio of 
$B \to X_s \gamma$ \cite{cleo} :
$2.0 \times 10^{-4} < {\cal B}(B \to X_s\gamma) < 4.5 \times 10^{-4}$ ;
(i) $\phi_{\mu} \leq 10^{-2}$ and $ 0 \leq \phi_{A} \leq 2 \pi$,
(ii) As in the case of no SUSY CP violating phase, 
$C_7 $ and $C_8$ have large SUSY contributions, however, those to $C_9$ and
$C_{10}$ are small. 
(iii) CP asymmetry of $B \to X_s \gamma$ : 
$A_{CP}(B \to X_s \gamma) \leq 2 \%$ with EDM constraints, however
when we consider EDM cancellation in one loop level \cite{kane},
it can be reached upto $7\%$.
(iv) In $B \to X_s \ell^{+}\ell^{-}$ decay, for small $tan\beta$ value,
$Im(C_7/C_7^{sm}) \simeq 0$ since $Im(A_t)$ becomes small, however,   
for large $tan\beta$ value, since chargino and stop loop effect becomes
dominated in $C_7$, $C_7 \simeq \pm C_7^{sm}$. So branching ratio of
$B\to X_s\ell^{+}\ell^{-}$ can be enchanced when $C_7 \simeq -C_7^{sm}$.
(v) The allowed domain of $C_8/C_8^{sm}$ can be extracted from the 
$B \to X_s g \gamma$ contribution into $B \to X_s \gamma$, 
shown in Figure 1.

When we investigate the new physics effect through magnetic gluon penguin
contributions in the exclusive B meson decays, we find that a possible
large direct CP violation can be observed in pure penguin modes, specially
$B^{\pm} \to \phi K^{\pm}$ and $K^{0}\pi^{\pm}$.

The CP asymmetry is defined as :
\begin{eqnarray}
A_{CP} &=& {\Gamma(B^{-} \to f) - \Gamma(B^{+} \to \bar{f}) \over
\Gamma(B^{-} \to f) + \Gamma(B^{+} \to \bar{f})}
\nonumber 
\end{eqnarray}
For instance, the amplitude of $B^{+} \to \phi K^{+}$ decay is :
\begin{eqnarray}
A(B^{+} \to \phi K^{+}) &=&
-{G_F \over \sqrt{2}} V_{tb}V^{*}_{ts} \left[ a_3 + a_4 + a_5
-{1 \over 2}(a_7 + a_9 + a_{10}) + a_{8G} \right] \,\, M^{(BK,\phi)} 
\nonumber \\ 
M^{(BK,\phi)} &=& <\phi|(\bar{s}s)_{V-A}|0><K|(\bar{b}s)_{V-A}|B>
= 2 f_{\phi} \,\, m_{\phi} \,\, [\epsilon \cdot P_B] F_1^{BK}(m^2_{\phi})
\nonumber
\end{eqnarray} 
where 
\begin{eqnarray}
a_{8G} &=& {\alpha_s \over 4 \pi} {m_b^2 \over q^2} C_{8G} 
{N_c^2 -1 \over N_c^2} \,\,
S_{\phi K} \, \cdot \, e^{i \sigma}  \nonumber \\
C_{8G} &=& C_{8G}^{new} + C_{8G}^{sm} = C_{8G}^{sm} \, 
\cdot \,R \, \cdot \, e^{i\theta} \nonumber
\end{eqnarray}
where $S_{\phi K} = - 0.49$, $\sigma$ is the strong phase difference 
between $O_{8G}$ and $O_{1-10}$, $\theta = \delta_{new} - \delta_{sm}$ is
the electroweak phase difference between new physics and SM, and 
$R = |C_{8G}/C_{8G}^{sm}|$.
In our analysis we use the factroization approach including
non-factorizable contributions into $N_c^{eff}$ and strong phases via
Bander-Silverman-Soni mechanism \cite{bss} into $a_{i}$ and $C_{i}^{eff}$.
We use $(N_c^{eff})_{LL} = 2.0$ for $O_{1,2,3,4,9,10}$ and
$(N_c^{eff})_{LR} = 6.0$ for $O_{5,6,7,8}$ 
as like as H.Y. Cheng et al.\cite{hycheng}.
As shown in Figure 2, we have large direct CP asymmetries which is induced
by new physics contributions :
in $B^{\pm} \to \phi K^{\pm}$ decay, we have 
$0.1 \times 10^{-5} \leq {\cal B}(B^{\pm} \to \phi K^{\pm}) \leq 0.75 \times
10^{-5}$ and $|A_{CP}| \leq 15 \%$ 
with EDM constrained condition, however, without EDM constraints,
$|A_{CP}| \leq 30 \%$.
For $B^{\pm} \to K^0 \pi^{\pm}$ decay, we get the branching ratio upto
$22.5 \times 10^{-6}$ which is well agreed 
with present experimental data \cite{cleo2}:
${\cal B}(B^{\pm} \to K^o \pi^{\pm}) = (18.2^{+4.6}_{-4.0} \pm 1.6) 
\times 10^{-6}$ and $|A_{CP}| \leq 20 \%$.  

In conclusion we have given example of decay modes which can allow
an early detection of new physics effects in the minimal supergravity model.
\section*{Acknowledgments}
The author wishes to thank T. Goto, T.Nihei, Y. Okada and Y.Shimizu
for collaboration on some of the work presented here.
Y.Y.K. would like to thank M. Kobayashi and H.Y Cheng for their
hospitality.

\begin{nfigure}{tb}
\centerline{\epsfxsize=0.6\textwidth \epsffile{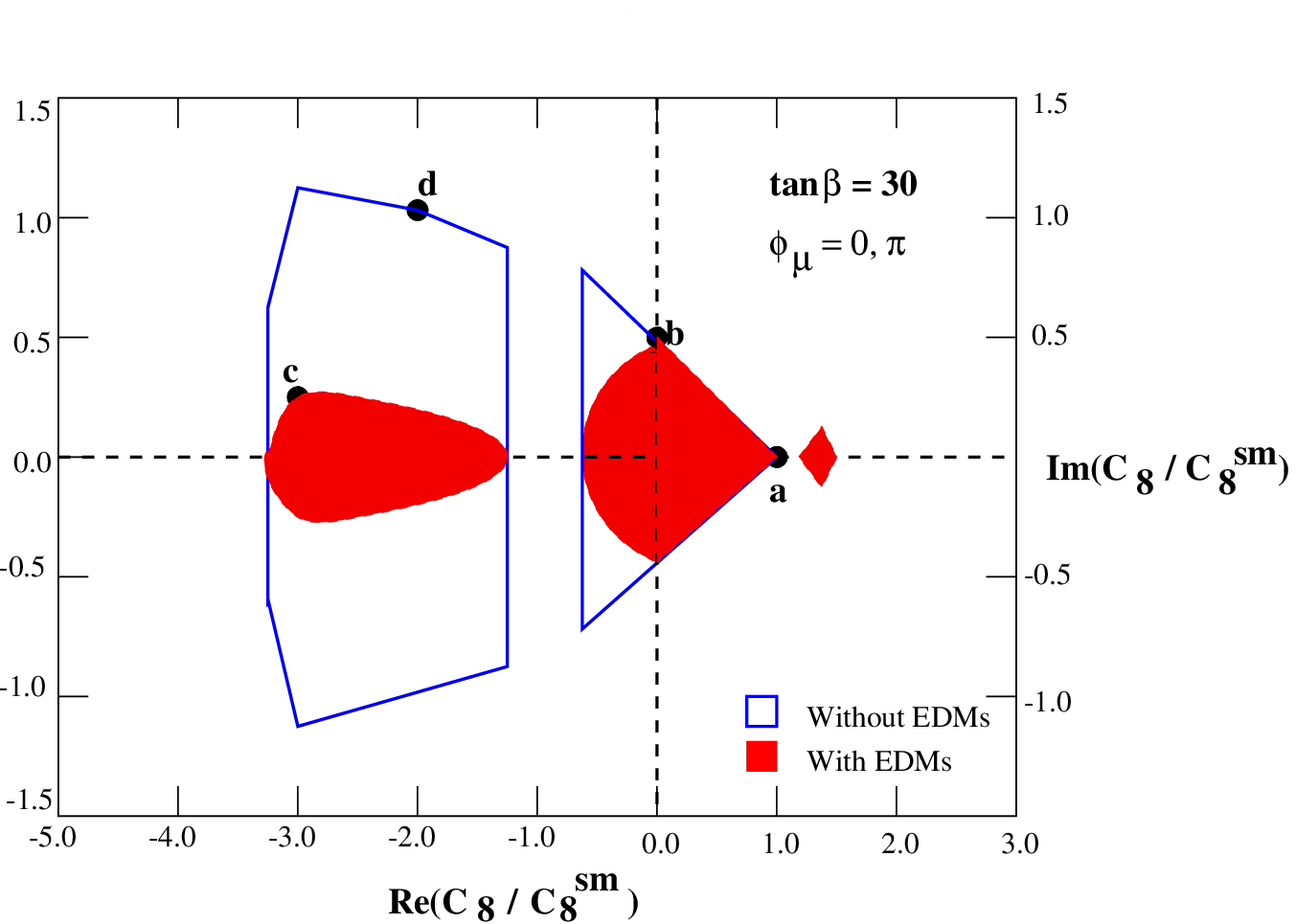}}
\caption{The allowded region of $C_8/C_8^{sm}$ at the bottom
mass scale with $tan\beta = 30$.  \label{fig:c8G}}
\end{nfigure}
\begin{nfigure}{tb}
\centerline{\epsfxsize=0.5\textwidth \epsffile{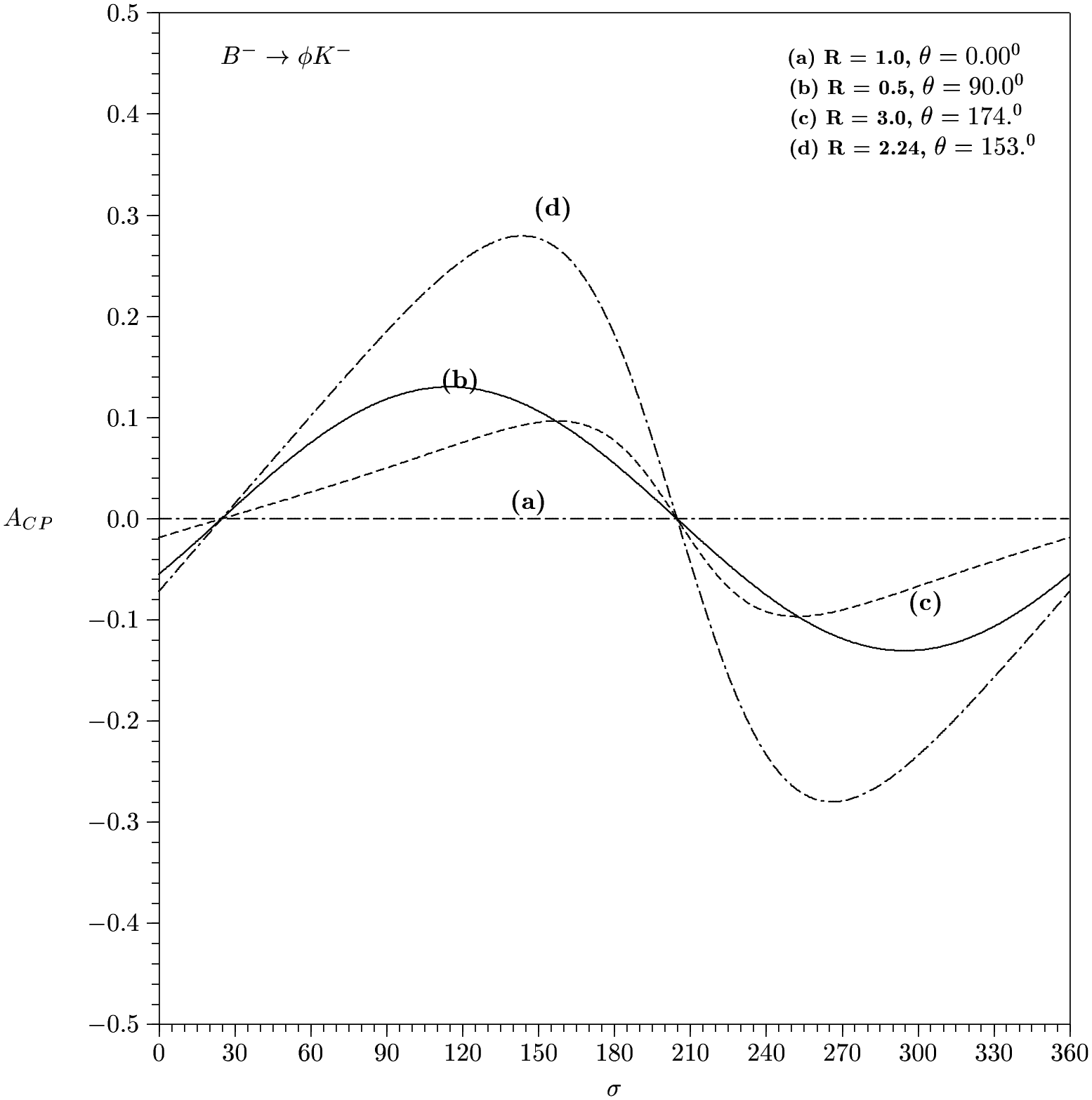}}
\caption{CP asymmetry at four different points: 
(a) prediction of the Standard Model,
(b) the point with pure imaginary of $C_8$,
(c) the point with maximum distance from origin,
(d) an example point without EDM constaint. \label{fig:AcpKphi}}
\end{nfigure}

\end{document}